\documentclass[preprintnumbers, showpacs, amsmath, aps, twocolumn]{revtex4}
\usepackage[dvips]{graphicx}
\usepackage{dcolumn}
\usepackage{bm}
\begin{document}

\title{Ferromagnetic resonance in systems with competing uniaxial and cubic anisotropies}
%\subtitle{FMR for mixed uniaxial and cubic anisotropies}
\author{Hamid Kachkachi}
\email{hamid.kachkachi@physique.uvsq.fr}
\affiliation{Groupe d'Etude de la Mati\`ere Condens\'ee, Universit\'e de Versailles St. Quentin, CNRS UMR8635, 45 av. des Etats-Unis, 78035 Versailles, France}
\author{David S. Schmool}
\email{dschmool@fc.up.pt}
\affiliation{Depto. de Fisica and IFIMUP, Universidade do Porto, Rua Campo Alegre 687, 4169 007 Porto, Portugal}

\date{\today}

\begin{abstract}
We develop a model for ferromagnetic resonance in systems with competing uniaxial and cubic anisotropies.
This model applies to i) magnetic materials with both uniaxial and cubic anisotropies, and ii) magnetic nanoparticles with effective core and surface anisotropies.
We numerically compute the resonance frequency as a function of the field and the resonance field as a function of the direction of the applied field for an arbitrary ratio of cubic-to-uniaxial anisotropy. 
We also provide some approximate analytical expressions in the case of weak cubic anisotropy.
We propose a method that uses these expressions for estimating the uniaxial and cubic anisotropy constants, and for determining the relative orientation of the cubic anisotropy axes with respect to the crystal principle axes. 
This method is applicable to the analysis of experimental data of resonance type measurements for which we give a worked example of an iron thin film with mixed anisotropy.
\end{abstract}
\maketitle

\section{Introduction}
The dynamics of magnetic nanoparticles is an area of intense theoretical and experimental investigation.
From the technological point of view, one of the reasons for such a great interest partly stems from the growing demands on the magnetic recording industry.
However, small nanoparticles, used for high density storage, become superparamagnetic even at low temperature because of the thermal instability of their magnetisation. 
Controlling this effect, in view of room temperature applications, requires an understanding of the magnetisation dynamics at the nanosecond time scale, taking into account the various material properties.
The magnetic properties of a fine nanoparticle, as compared to the bulk material, are dramatically altered due to its reduced size and to the boundary effects which are induced by the modified atomic environment at its surface. Consequently, the nanoparticle turns out to be a many-body system whose magnetic state cannot, {\it a priori}, be faithfully described by a macroscopic approach.
On the other hand, from the theoretical point of view, investigating the dynamic properties of a many-spin particle is a real challenge because one is faced with the inherent difficulties related to the analysis of the energy potential. This analysis is unavoidable since it is a crucial step in the study of ferromagnetic resonance (FMR) and the dynamic reversal of the particle's magnetisation.
Nevertheless, there exist some regimes of the physical parameters where the macro-spin approach may still be used if appropriately extended. 
For example, in Refs.~\cite{garkac03prl, kacbon06prb, yanesetal07prb} it has been shown, analytically as well as numerically, that when the surface anisotropy constant is much smaller than the exchange coupling, the surface anisotropy contribution to the particle's energy is of $4^{\mathrm{th}}$-order in the net magnetisation components and $2^{\mathrm{nd}}$-order in the surface anisotropy constant. This means that the behaviour of a many-spin particle with uniaxial anisotropy in the core and relatively weak surface anisotropy (transverse or N\'eel) can be modelled by that of a macro-spin system whose effective energy contains an additional cubic-anisotropy term.
On the other hand, magnetocrystalline anisotropy of $4^\mathrm{th}$-order naturally arises in magnetic materials (e.g., iron and YIG) and adds to the $2^\mathrm{nd}$-order contribution, though with an order of magnitude smaller [see Ref.~\cite{jametetal01prl}]. 
In iron magnetic multilayers and thin films we also observe a mixing of uniaxial and cubic anisotropies which arise from their growth on particular substrates, notably GaAs $(001)$ \cite{heicoc93ap}.

In Ref.~\cite{jametetal01prl} the effective macro-spin approach was used to interpret the $\mu$-SQUID measurements of the $3D$ switching field (or Stoner-Wohlfarth astroid) for a cobalt particle of $3$ nm diameter (containing $1500$ atoms).
The authors concluded that the cluster-matrix interface may be responsible for the main contribution to the magnetic anisotropy of the nanocluster.
This gives a further indication that the macro-spin approach with an effective potential energy provides a more reasonable approximation to the initial many-spin particle than the macro-spin Stoner-Wohlfarth model. This may thus be used to investigate, though in a phenomenological manner, the dynamics of the particle, and in particular the FMR characteristics.

Accordingly, we first consider the general situation of either i) a single magnetic moment, regarded as a macro-spin representation of a nanoparticle, with an effective potential containing both a $2^\mathrm{nd}$- and a $4^\mathrm{th}$-order anisotropy terms, including the applied magnetic field, or ii) a magnetic material with competing magnetic anisotropies (uniaxial and cubic).
We compute the resonance field and frequency as functions of the intensity of the cubic-anisotropy contribution and the static magnetic field (with varying direction and magnitude).
In addition, we also consider the possibility of having the cubic-anisotropy axes rotated to some angle with respect to the axes of the crystal lattice, while the uniaxial anisotropy easy axis is maintained parallel to the $z$ axis throughout.

This work is organised as follows: we first outline the basic framework of the free energy and parametrise it into dimensionless quantities which is convenient for the numerical calculations. 
The free energy, containing a uniaxial and cubic anisotropy terms together with the Zeeman contribution, is then used to evaluate the ferromagnetic resonance conditions. 
We perform two types of calculation: i) the resonance field as a function of the direction of the applied field, and ii) the full frequency spectrum as a function of the applied field.  In each case we have varied the relative strengths of the anisotropy constants to demonstrate how the resonance condition changes with the growing influence of the cubic-anisotropy contribution.
After discussing the results for the general case, we consider the specific case of a cobalt nanoparticle as a model system, with the parameters of Ref.~\cite{jametetal01prl}, to estimate the range of the resonance frequency and field. 
We also estimate the anisotropy constants in an iron thin film with both uniaxial and cubic anisotropies by fitting experimental data.
\section{Energy and physical parameters}
\subsection{Basic expressions and notation}
We define the system's net magnetic moment as $\mathbf{m}=\mu _{s}\,\mathbf{s}$ with $\mu_s=m_{v}V$, where $m_{v}$ is the magnetic moment density, $\mathbf{s}$ ($|\mathbf{s}|=1$) is the unit vector in the direction of $\mathbf{m}$.
The energy of $\mathbf{m}$ reads
\begin{equation}\label{InitialEnergy}
E=-\mu _{s}H\,\left( \mathbf{s}\cdot \mathbf{e}_{h}\right) -K_{2}V\,(\mathbf{
s}\cdot \mathbf{n})^{2}+\frac{K_4 V}{2}\,(s_{x'}^{4}+s_{y'}^{4}+s_{z'}^{4}),
\end{equation}
where $\mathbf{n}$ is the uniaxial anisotropy easy axis ($K_{2}>0$) and $\mathbf{e}_{h}$ the
unit vector along the applied field.
In Eq. (\ref{InitialEnergy}) we have used a different form for the cubic anisotropy from that often found in the literature, i.e., $-K_4V\left(
s_{x}^{2}s_{y}^{2}+s_{y}^{2}s_{z}^{2}+s_{z}^{2}s_{x}^{2}\right)$, with $K_4>0$. These forms are related through the identity $1=\left( s_{x}^{2}+s_{y}^{2}+s_{z}^{2}\right)
^{2}=s_{x}^{4}+s_{y}^{4}+s_{z}^{4}+2\left(s_{x}^{2}s_{y}^{2}+s_{y}^{2}s_{z}^{2}+s_{z}^{2}s_{x}^{2}\right)$, where the first equality is due to $|\mathbf{s}|=1$.
The ($x'y'z'$) coordinate system is deduced from ($xyz$) by a rotation of $\psi$ around the $z$ axis, i.e., $s_{\alpha'} = \sum_{\beta=x,y,z}R^{\alpha\beta}\,s_\beta$, where $R^{\alpha\beta}$ are the matrix elements of the corresponding rotation.

The magnetic moment $\mathbf{m}$ experiences the effective field defined by $\mathbf{H}_{\mathrm{eff}} \equiv -\frac{1}{\mu _{s}}{\delta E}/{\delta\mathbf{s}}$ and which is normalized with respect to the second-order anisotropy field 
\begin{equation}\label{AnisotropyFieldMax}
H_2\equiv \frac{2K_{2}V}{\mu _{s}},  
\end{equation}
leading to the dimensionless field vector 
\begin{equation}\label{DimensionlessEffectiveField}
\mathbf{h}_{\mathrm{eff}}=h\,\mathbf{e}_{h}+k_{2}\,(\mathbf{s}\cdot \mathbf{n})\mathbf{n}-\,\zeta\sum\limits_{\alpha,\beta=x,y,z}\,s_{\beta'}^{3}R^{\beta\alpha}\mathbf{e}_{\alpha},
\end{equation}
where 
\begin{equation}\label{ReducedHK4}
h\equiv \frac{H}{H_2},\quad \zeta \equiv \frac{K_4}{K_2}, 
\end{equation}
and the parameter $k_{2}=0,1$ is inserted to allow us to switch on or off the uniaxial anisotropy. 
Using these dimensionless quantities, the energy (\ref{InitialEnergy}) becomes 
\begin{equation}\label{DimensionlessEnergy}
\mathcal{E}\equiv \frac{E}{2K_{2}V}=-h\,\mathbf{s}\cdot \mathbf{e}_{h}-\frac{%
k_{2}}{2}\,(\mathbf{s}\cdot \mathbf{n})^{2}+\frac{\zeta }{4}\sum\limits_{\alpha =x,y,z}s_{\alpha'}^{4}.  
\end{equation}

Using the parametrisation $\mathbf{s}(\theta ,\varphi )$ and $\mathbf{e}_{h}(\theta _{h},\varphi _{h})$ the (dimensionless) FMR equation reads [see e.g., Ref.~\cite{gurmel96crcpress, farle98rpp} and references therein]
\begin{equation}\label{FMREqUCA}
\omega _{s}^{2}=\frac{1}{\sin ^{2}\theta }\left[ \left( \partial _{\varphi
}^{2}\mathcal{E}\right) \left( \partial _{\theta }^{2}\mathcal{E}\right) -\left( \partial _{\theta \varphi }^{2}\mathcal{E}
\right) ^{2}\right] ,  
\end{equation}
with the right-hand side being evaluated at the equilibrium state ${\bf s}_0(\theta_0, \varphi_0)$. 
We have also introduced the reduced angular frequency 
\begin{equation}
\omega _{s}\equiv \omega \,\frac{\mu _{s}}{2\gamma K_{2}V}=\omega \tau _{s},
\label{ReducedAngularFrequency}
\end{equation}
with 
\begin{equation}\label{ScalingTime}
\tau _{s}\equiv \frac{\mu _{s}}{2\gamma K_{2}V} =  \frac{1}{\gamma H_2}
\end{equation}
being the scaling time. So, the reduced frequency reads
\begin{equation*}
\nu _{s}\equiv \frac{\omega _{s}}{2\pi }=\frac{\omega \tau _{s}}{2\pi }=\nu
\tau _{s}.  
\end{equation*}

As mentioned in the introduction, the appearance of the $4^\mathrm{th}$-order term in Eq.~(\ref{DimensionlessEnergy}) may be of two origins: i) the natural magnetocrystalline cubic-anisotropy contribution that adds to the uniaxial anisotropy in real materials \cite{gurmel96crcpress, farle98rpp, jametetal01prl}, or ii) as has been shown in Refs.~\onlinecite{garkac03prl, kacbon06prb, yanesetal07prb}, the contribution induced by the surface anisotropy in a nanoparticle cut from a cubic lattice.

While the first situation is quite common in magnetic materials, and in particular in thin films, the second case deserves further explanation and a review of the recent results on which the underlying assumption is based. This is done in the following section
\subsection{Effective energy of a nanoparticle}
Investigating the dynamics of a nanoparticle taking account of its intrinsic properties, such as its size and shape, crystal structure and surface anisotropy, would require the use of an ``atomistic'' approach with the atomic magnetic moment as its building block.
However, within this approach one is faced with complex many-body aspects with the inherent difficulties related with analysing the energyscape (location of the minima, maxima, and saddle points of the energy potential). This analysis is unavoidable since it is a crucial step in the calculation of the relaxation time and thereby in the study of the magnetization stability against thermally-activated reversal. 
In view of such difficulties, one may then ask the question as to whether there exists an intermediate approach with the relative simplicity of the macroscopic approach and richness of the many-spin approach, namely a macroscopic model which captures some of the intrinsic features of the magnetic nanoparticle.
In Refs. \onlinecite{garkac03prl, kacbon06prb, yanesetal07prb} analytical as well as numerical calculations showed that a many-spin particle, cut from a cubic lattice, when its surface anisotropy is small with respect to the exchange coupling, i.e., when its magnetic state is not far from the collinear state, may indeed be modeled by an effective one-spin problem (EOSP), i.e., a single macroscopic magnetic moment $\mathbf{m}$ representing the net magnetic moment of the many-spin particle. The energy of this EOSP (normalized to $J{\cal N}$, where $J$ is the substance bulk exchange coupling and ${\cal N}$ the total number of spins in the cluster) may be written as
\begin{equation}
\mathcal{E}_{\mathrm{EOSP}}=\mathcal{E}_c + \mathcal{E}_1^{(0)} + \mathcal{E}_2^{(1)} + \mathcal{E}_2^{(0)}.
\label{eq:EOSPEnergy}
\end{equation}
where i) $\mathcal{E}_c$  is the pure core anisotropy contribution that may be uniaxial, cubic, bi-axial, etc. ii) $\mathcal{E}_2^{(0)}$ is the pure surface contribution that stems from surface anisotropy; it is quadratic in the single-site surface anisotropy constant $K_s$ and quartic in the components of $\mathbf{m}$; it is also proportional to a surface integral that depends on the size, shape, and crystal structure of the initial many-spin particle [see Eq.~(\ref{Keff}) below]. iii)  the contribution  $\mathcal{E}_{1}^{(0)}$ is induced by elongation (or shape anisotropy); this term is quadratic in the components of $\mathbf{m}$ and linear in $K_s$. iv) $\mathcal{E}_{2}^{(1)}$ arises from a competition between the surface anisotropy which induces spin disorder that tends to propagate deep into the particle, and the core anisotropy that tends to expel such spin-noncollinearities out to the particle's border. This core-surface mixing contribution is linear in the core anisotropy constant $K_c$, quadaratic  in  $K_s$,  and its dependence on $\mathbf{m}$ is given by a function that mixes quadratic and quartic anisotropies. However, the contribution  $\mathcal{E}_{2}^{(1)}$ is only relevant if  $\left(K_c/J\right) \gtrsim (K_s/J)^2$.

Collecting all contributions, a many-spin particle satisfying the above-mentioned conditions, may be described by a macroscopic magnetic moment $\mathbf{m}$, representing the net magnetic moment of the particle, whose (effective) energy may be written as in Eq.~(\ref{DimensionlessEnergy}) with $k_2,\zeta$ being regarded as the effective uniaxial and cubic anisotropy constants, respectively.
It is then important to note that the magnitude and sign of these constants depend on the intrinsic properties of the initial many-spin particle, such as the crystal structural, size and shape, and physical parameters such as the single-site surface anisotropy (in intensity and model). However, we should emphasize that these results  hold for cubic crystal lattices and quadratic surface anisotropy models, such as N\'eel's or transverse.

For further reference, we note that in Ref.~\onlinecite{garkac03prl} an analytical expression was given for the effective constant $K_\mathrm{eff}$ of the surface-induced cubic-anisotropy term $\mathcal{E}_2^{(0)}$, when the core anisotropy is absent, that is
\begin{equation}\label{Keff}
K_\mathrm{eff} = \kappa \frac{K_s^2}{z J},
\end{equation}
where $z$ is the coordination number and $\kappa$ a surface integral that depends on the underlying lattice, shape, and size of the particle and also on the surface-anisotropy model. For a spherical particle (of $\sim 1500$ spins) cut from a simple cubic lattice and with N\'eel's surface anisotropy, $\kappa\simeq 0.53465$.

Finally, we note that the shape anisotropy is included in the uniaxial anisotropy contribution. We also assume that in ellipsoidal particles the magneto-crystalline easy axis is parallel to the direction of the major axis.
\section{\label{sec:results}Results and discussion}
Before we discuss our results, some remarks are in order concerning the general numerical method used here.
In order to obtain, for instance, the resonance field from Eq.~(\ref{FMREqUCA}), for a given angular frequency $\omega_{s}$, one has to find the equilibrium state of the system for a given set of the physical parameters $k_{2},\zeta ,h,\theta_h,\varphi_h$.
However, as it is not possible to obtain in general analytical expressions for the equilibrium state,  or the global minimum of the energy (\ref{DimensionlessEnergy}), we resort to numerical approaches. Accordingly, since we only need the absolute minimum, an adequate numerical method is provided by the standard Metropolis algorithm with random increments, which is a global method. 
Next, once the global minimum has been found, we proceed with a fine search by solving the Landau-Lifshitz equation with weak damping using the Runge-Kutta method.
%
%=========================================================================
\begin{figure*}[floatfix]
\includegraphics[width=5.5cm, angle=-90]{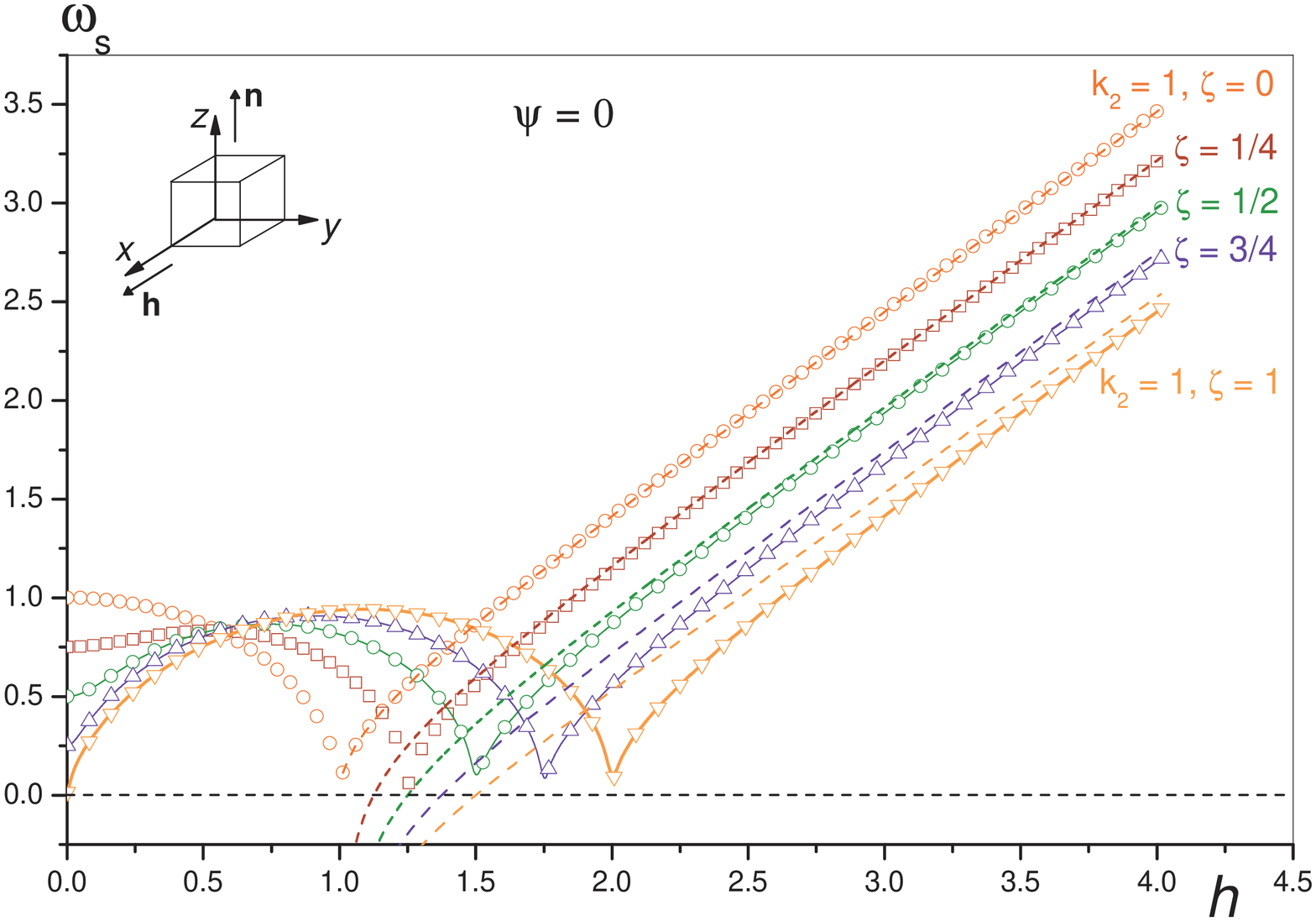}
\includegraphics[width=5.5cm, angle=-90]{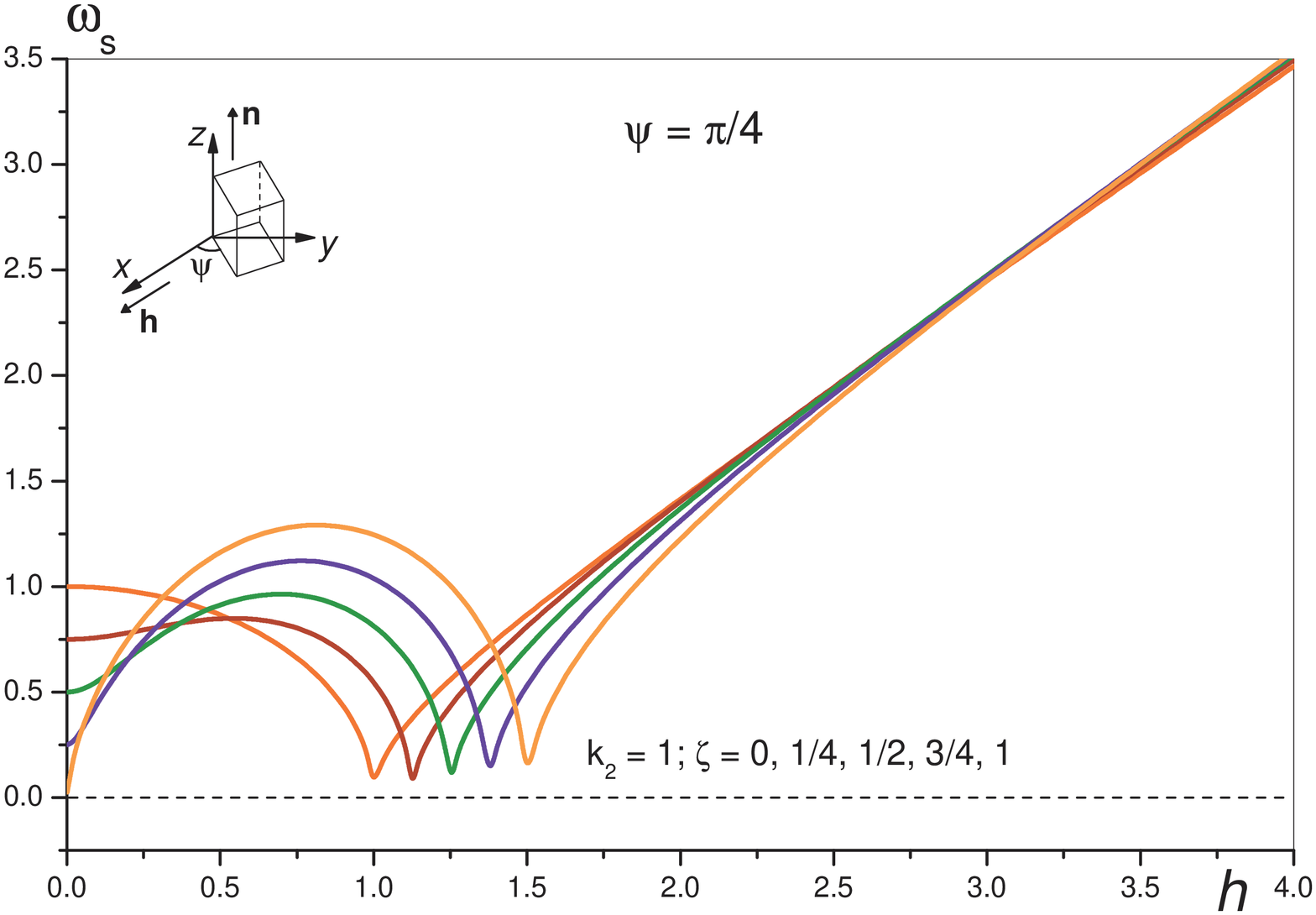}
\caption{Resonance frequency as a function of the applied field for various values of $\zeta$ with the cubic-anisotropy axes making an angle $\psi=0,\pi/4$ with respect to the $(x,y,z)$ frame.}
\label{OmhUCA_psi045}
\end{figure*}
%===========================================================================
%

In Fig.~\ref{OmhUCA_psi045} we plot the resonance frequency as a function of the applied field for various values of $\zeta$ with the cubic-anisotropy axes making an angle $\psi=0,\pi/4$ with respect to the $(x,y,z)$ frame.
We now discuss some features common to both cases of $\psi=0$ and $\pi/4$: 
i) In the case of zero field, using the FMR equation (\ref{FMREqUCA}), or the effective field (\ref{DimensionlessEffectiveField}), with small $\zeta$, the global minimum is predominantly determined by the uniaxial anisotropy, such that $\omega_s\simeq k_2 - \zeta$. This explains the decrease of $\omega_s$ with $\zeta$ at zero field, since the uniaxial- and cubic- anisotropy terms contribute with opposite signs [see Eq.~(\ref{DimensionlessEffectiveField})]. 
Obviously, this behaviour does not change when the cubic-anisotropy axes are rotated by some angle around the $z$ axis, as can be seen in Fig.~\ref{OmhUCA_psi045} (right), due to the rotation symmetry in the equator ($xy$ plane) when $h=0$.
ii) In strong fields, $\theta_0\sim\theta_h,\varphi_0\sim\varphi_h$ and we can expand the FMR equation (\ref{FMREqUCA}) in $(\zeta/h)$, assuming weak cubic anisotropy, to obtain the asymptotic behaviour (dashed lines in Fig.~\ref{OmhUCA_psi045} (left)). 
This leads to
\begin{eqnarray}\label{asymptotes}
\omega_s &\simeq& h r - \frac{h}{2}\left(\frac{a}{r} + b r\right)\frac{\zeta}{h}, \\ \nonumber
r &=& \sqrt{1 - \frac{k_2}{h}}, \quad a = \cos^4\psi + \sin^4\psi, \quad b = \cos4\psi.
\end{eqnarray}
The coefficient of the term in $\zeta/h$ is $(r+1/r)/2$ for $\psi=0$ and $(r-1/2r)/2$ for $\psi=\pi/4$.
Hence, for $\psi=\pi/4$ the effect of the correction term in $\zeta/h$ is negligible. Indeed, as can be seen in Fig.~\ref{OmhUCA_psi045} (right), the high-field asymptote is almost independent of $\zeta$ and is approximately given by the asymptote for the uniaxial anisotropy, i.e., by Eq.~(\ref{asymptotes}) with $\zeta=0$. 

A point on the curves $\omega_s(h)$ that is of special importance in practice, is that defined by $(h=h_c,\omega_s=0)$ [see discussion below].
In Fig.~\ref{OmhUCA_psi045} the critical field at which $\omega_s$ vanishes is given by
\begin{equation}\label{criticalfield}
h_c = k_2 + \zeta\left[\cos^4\psi + \sin^4\psi\right]. 
\end{equation}
In particular, we have $h_c=k_2+\zeta$ in Fig.~\ref{OmhUCA_psi045} (left) and $h_c=k_2+\zeta/2$ in Fig.~\ref{OmhUCA_psi045} (right).
This, together with the analysis for $h=0$ and the asymptote (\ref{asymptotes}) for strong fields, shows that the effect of the $\psi$ rotation is to reduce the influence of cubic anisotropy.
In the case of pure cubic anisotropy ($k_2=0$) the curves of $\omega_s$ versus $h$ cross the other curves with $k_2\neq 0$, and for the sake of clarity were not included in Fig.~\ref{OmhUCA_psi045}.
One should note that in fact the critical field $h_c$, at which $\omega_s$ tends to zero, is the field on the Stoner-Wohlfarth curve (or astroid) at which the metastable minimum merges with the saddle point and loses its local stability.
Indeed, from Eq.~(\ref{FMREqUCA}) we see that the condition $\omega_s=0$ is just the definition of this inflection point. 

The above results can be used to interpret FMR measurements with sweeping frequency as obtained, for example, by the Network Analyser FMR (NA-FMR) technique \cite{mosetal06jmmm}, Brillouin Light Scattering (BLS) \cite{cochran94springer} and Pump-Probe Microscopy (PPM) \cite{hiebertetal79prl}. More precisely, one can extract the anisotropy constants $K_2,K_4$, and also the rotation angle $\psi$. 
For this, we can use the three independent conditions provided by i) $\omega_s$ at zero field which is $k_2-\zeta$, ii) the high-field asymptote (\ref{asymptotes}), and iii) the critical point $h_c$ given by (\ref{criticalfield}) at which $\omega_s$ tends to zero.
Therefore, for a given material with given uniaxial and cubic anisotropies, in which the cubic-anisotropy axes are at some arbitrary azimuthal angle $\psi$ with respect to the crystal axes, we can uniquely determine the three parameters ($K_2,K_4,\psi$).
It should be noted, however, that these measurements must be made with the field applied along one of the hard axes in order to obtain the three above conditions. For any other orientation of the applied field we lose condition iii), because the cusp is no longer well defined. 
In the general situation, where the uniaxial-anisotropy axis is tilted at some angle with respect to the $z$ axis ($[001]$ axis), the above three conditions should be rederived.

%=========================================================================
\begin{figure*}[floatfix]
\includegraphics[width=5.5cm, angle=-90]{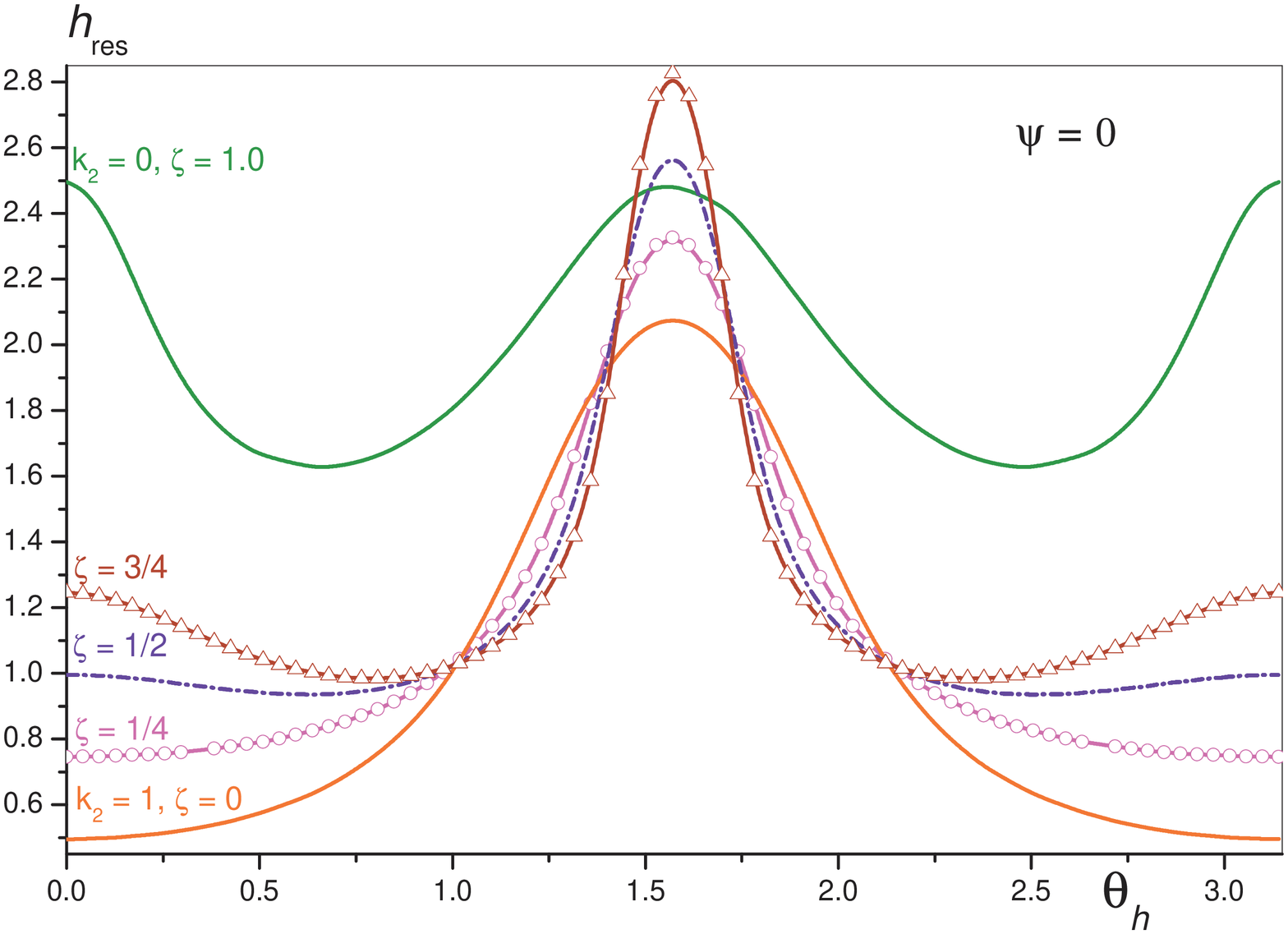}
\includegraphics[width=5.5cm, angle=-90]{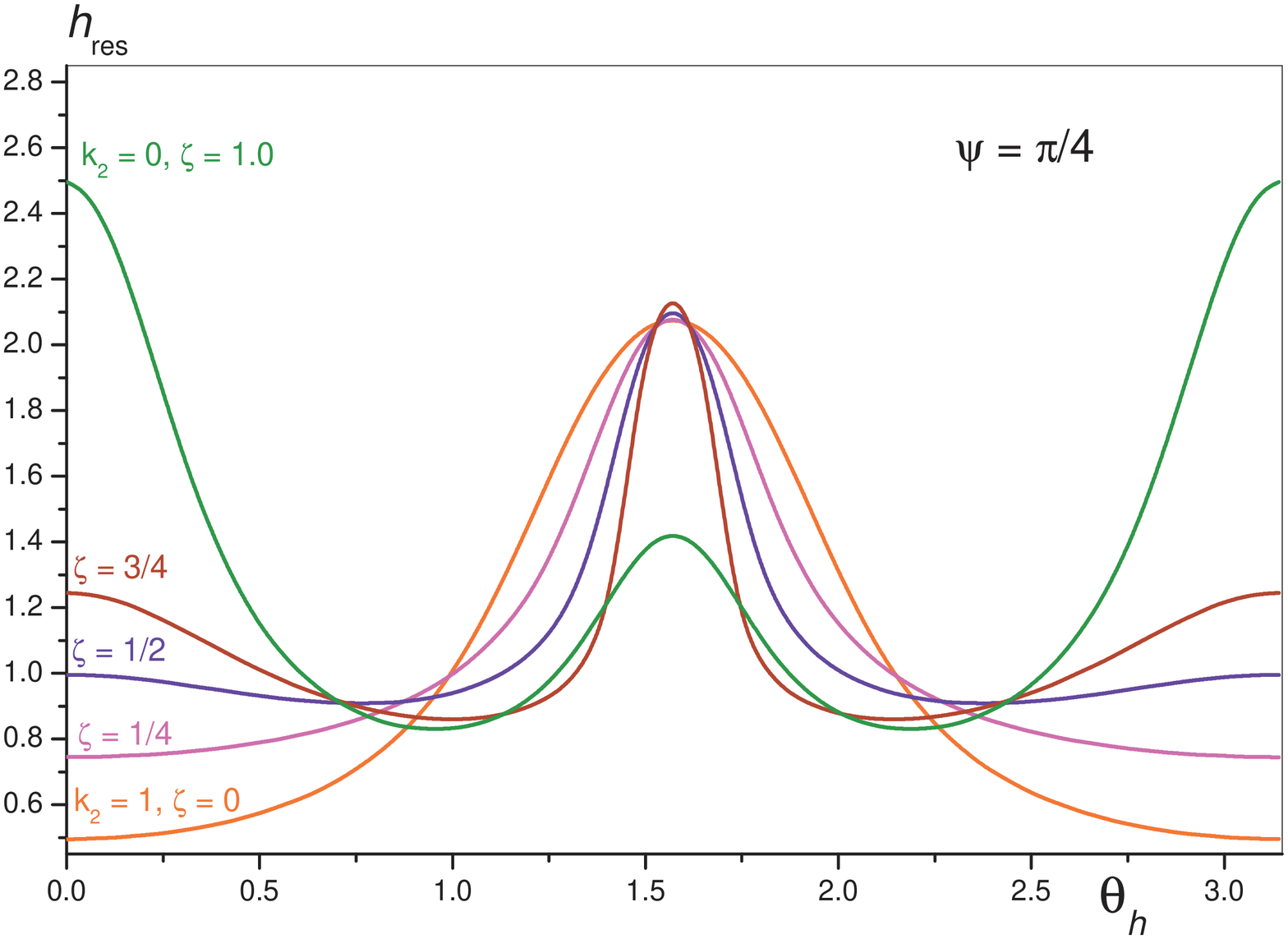}
\caption{Resonance field versus the direction of the applied field for various values of $\zeta$. The field is applied in the $xz$ plane. The frequency is set to $\omega_s=1.5$, which corresponds to $13.5$ GHz. Also shown for comparison is the case of pure cubic anisotropy ($k_2=0, \zeta=1$). The cubic-anisotropy axes are at $\psi=0,\pi/4$ with respect to the $(x,y,z)$ frame. (The angle $\theta_h$ is measured in radians.)}
\label{HresTH_UCA_zeta_psi045}
\end{figure*}
%===========================================================================
%

Fig.~\ref{HresTH_UCA_zeta_psi045} shows plots of the resonance field versus the direction of the applied field for various values of $\zeta$ and for the cubic-anisotropy axes parallel and at $\psi=\pi/4$ with respect to the $(x,y,z)$ frame. 
These plots are for the fixed frequency $\omega_s=1.5$, which corresponds to $13.5$ GHz in physical units.
In effect, this corresponds to cutting a horizontal line (at $\omega=1.5$) in the frequency-field curves (Fig.~\ref{OmhUCA_psi045}).
The curves with $k_2=0,\zeta=1$ correspond to the case of pure cubic anisotropy and are included only for comparative purposes.
At $\theta_h=0$, we have $h_\mathrm{res}\simeq \omega - (k_2 - \zeta)$. 
In strong fields, we can compute the resonance field, in principle, by replacing in Eq.~(\ref{asymptotes}) $\omega_s$ by a fixed frequency and solving for $h$. This should lead to an approximate expression, albeit somewhat cumbersome, for the resonance field at $\theta_h=\pi/2$.
For this orientation, we see that the difference induced by the cubic anisotropy ($\zeta$) is suppressed by the $\psi$ rotation.
As for the frequency plots (Fig.~\ref{OmhUCA_psi045}), comparing the results in Fig.~\ref{HresTH_UCA_zeta_psi045} we see that the effect of the $\psi$ rotation is again to reduce the influence of cubic anisotropy.
Angular FMR studies in thin films also exhibit the angular variations indicated in Fig.~\ref{HresTH_UCA_zeta_psi045} \cite{heinrich94springer}.

By way of illustration and to give some orders of magnitude of the various constants of the resonance frequency and field, we first consider the example of a nanoparticle, and in particular a (faceted) truncated octahedral cobalt particle (of $3$ nm in diameter) as obtained experimentally in  Ref.~\onlinecite{jametetal01prl}.
In this reference, the $3D$ experimental switching field was fit to Eq. (\ref{DimensionlessEnergy}) with an additional (relatively weak) anisotropy term along the hard-axis $y$. According to the estimations given in \cite{jametetal01prl}, $K_2\simeq 2.2\times 10^5$ J/m$^3$ and $K_4\simeq 0.1\times 10^5$ J/m$^3$, which in our case yields $\zeta = K_4/K_2\simeq 0.05$. 
This implies that the effective anisotropy of this nanoparticle is mainly uniaxial.
On the other hand, using Eq.~(\ref{Keff}) and the parameters given thereafter, we estimate the surface anisotropy constant as $K_s\simeq 10^{-22}$ J/atom (or $0.1$ erg/cm$^2$), for the above mentioned cobalt particle of about $1500$ atoms.
Similar values have been quoted by several authors using Neutron (quasi)-Inelastic Scattering \cite{gazetal97epl} and FMR \cite{shilovetal99prb} on cobalt particles.
>From Eq.~(\ref{ScalingTime}), using $\mu_s\simeq 1.4\times 10^6$ A/m we obtain $\tau_{s}\simeq 1.8\times 10^{-11}$ s, and hence the angular frequency $\omega_s=1.0$ (in Fig.~\ref{OmhUCA_psi045}) corresponds to $\sim 9$ GHz, and $h_\mathrm{res}=1$ (in Fig.~\ref{HresTH_UCA_zeta_psi045}) to $H_\mathrm{res}=h_\mathrm{res}H_2\simeq 0.3$ T [see Eq.~(\ref{AnisotropyFieldMax})].

At present there are no experimental frequency-field data available on nanoparticle systems for comparison with theory. However, future work is expected to address this issue.
%
%=========================================================================
\begin{figure}[floatfix]
\includegraphics[width=8cm]{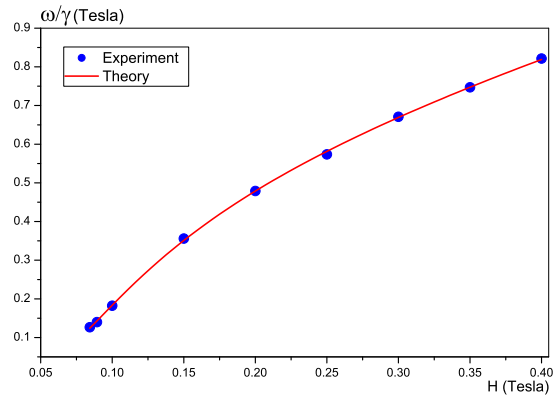}
\caption{Comparison of experiment and theory for the resonance frequency as a function of the applied field in a 40-monolayer iron thin film, as given in Ref.~\onlinecite{mosetal06jmmm}.}
\label{TheoryVSExperiments}
\end{figure}
%===========================================================================
%

As a second illustration we show the fitting procedure described earlier. For this we have used the frequency-field data from Ref.~\onlinecite{mosetal06jmmm} for a 40-monolayer iron film, obtained by NA-FMR, which exhibit mixed uniaxial and cubic anisotropies. 
This film is part of a bilayer system which may have some weak coupling. While the coupling itself is important in general, we have neglected its effect and only used this data as an illustration of the fitting procedure.
The results are shown in Fig.~\ref{TheoryVSExperiments} where full circles represent the experimental data while the line corresponds to the high-field expansion of Eq.~(\ref{asymptotes}), which we express as
\begin{eqnarray}\label{TheoryExpansion}
\omega_s &\simeq& h + a_0 + \dfrac{a_1}{h} + \dfrac{a_2}{h^2}, \\ \nonumber\\
a_0 &=& -\frac{1}{2}\left[1 + \left(2 - \frac{5}{2}\sin^2(2\psi)\right)\zeta\right], \nonumber\\
a_1 &=&  -\frac{1}{8}\left[1 + 3\sin^2(2\psi)\zeta\right], \nonumber\\
a_2 &=& -\frac{1}{16}\left[1 + \left(2 + \frac{1}{2}\sin^2(2\psi)\right)\zeta\right]. \nonumber
\end{eqnarray}
after setting $k_2=1$. This corresponds to the case with mixed uniaxial and cubic anisotropies.
The result of the best fit yields $H_2\simeq -0.175\mathrm{T}, \zeta \simeq 1.7$, and $\psi \simeq 17^{\circ}$.

It is seen that this approach leads to physically realistic estimates of the anisotropy parameters. It is also possible to consider the demagnetizing field contribution separately from the uniaxial anisotropy, and its value can be obtained experimentally as given in Ref.~\onlinecite{mosetal06jmmm}.
On the other hand, the present approach is applicable only at very low (or zero) temperature and as such overestimates the anisotropy constants for these measurements which were taken from experiments at room temperature.
Thermal effects on the resonance characteristics can be evaluated from the calculation of the transverse ac susceptibility as shown in Ref.~\onlinecite{raishl94acp}.
More generally, one also has to investigate the thermally-activated reversal of the magnetization and compute the relaxation rate of a magnetic moment with mixed uniaxial and cubic anisotropy and in a magnetic field applied at an arbitrary angle with respect to the uniaxial easy axis. This work is in progress \cite{dejardinetal07prep}.

On other hand, in the case of magnetic nanoparticle assemblies, one has to take account of the effect of inter-particle dipole-dipole interactions (DDI) on the static and dynamic properties.
In this context, and in the macroscopic approach, in Ref.~\onlinecite{kacaze05epjb} the static behaviour of the magnetization, as a function of temperature and applied field, was investigated taking account of anisotropy, DDI and also of the volume and anisotropy axes distributions. In the case of weak DDI, practical approximate analytical expressions were obtained by perturbation theory which provides a better approximation than the mean-field approach used in Ref.~\onlinecite{netzelmann90jap} [see also discussion in Ref.~\onlinecite{thomasetl95jmmm}].
These expressions involve many lattice tensors which account for the effect of the demagnetizing field and thus describe the change of magnetization in prolate and oblate particle systems. They also show how the magnetization deviates from the Langevin law in the presence of anisotropy and DDI.
Next, in Ref.~\cite{kacaze07prb} the effect of DDI on the dynamics of the assembly, and in particular on the zero-field-cooled magnetization was investigated, and an explanation was given for the change of behaviour of the maximum temperature as a function of the applied field.
It was shown that the transverse component of the DDI field creates new saddle points in the particle's energy and thereby increases the switching rate. In addition, the critical (or activation) volume that separates the superparamagnetic from blocked particles decreases upon increasing the particle concentration. It was found that this volume separates the low-field regime dominated by the blocked particles from the high-field regime dominated by the superparamagnetic ones. As such, as the concentration (or intensity of DDI) of the sample is increased, the low-field regime shrinks and eventually disappears.

In connection with the present work, the use of the developments in Refs.~\onlinecite{kacaze05epjb, kacaze07prb} is necessary in order to study the interplay between the effect of DDI and the intrinsic properties of nanoparticles modeled as an EOSP with the energy in Eq.~ (\ref{DimensionlessEnergy}), and the ensuing effects on the FMR characteristics.
\section{\label{sec:conc}Conclusion}
We have studied the ferromagnetic resonance of an effective magnetic moment in the general situation of an energy potential containing both uniaxial and cubic anisotropies.
In particular, we have computed the resonance frequency and field as functions of the applied field magnitude and direction.
These results can be used in interpreting the FMR measurements on magnetic materials which exhibit both forms of anisotropy.
We have provided a simple method for estimating the anisotropy constants (uniaxial and cubic) as well as evaluating the relative orientation of the cubic-anisotropy axes with respect to crystalline axes.
However, this model assumes that the uniaxial anisotropy axis is coincident with one of the  crystalline axes. The general case is currently being investigated and will be published in a future work.
The frequency-field curves, as illustrated here, can be obtained experimentally using techniques such as NA-FMR, BLS, and PPM.
On the other hand, the present FMR analysis may also be used to check whether this EOSP approach is a reasonable approximation to real magnetic systems and especially to nanoparticles.
However, it is not obvious how to quantitatively distinguish between the cubic anisotropy of  magnetocrystalline origin and the one that is induced by the surface contribution, since these two contributions turn out to be of the same order of magnitude.
Nevertheless, FMR measurements on well separated nanoparticles of a very small size and grown from a magnetic material of negligible cubic contribution could provide us with more precise data for this purpose.
One possible method would be to study the FMR spectra as a function of the particle's size. Indeed, changing the size should alter the relative contributions of surface and core anisotropies and thus indicate the origin of each.

We have used the model presented here to fit the experimental data from the high-field branch of the frequency-field curve of NA-FMR measurements on an iron thin film. We observe good agreement between theory and experiment which illustrates the application of the model.

In a recent work \cite{kacsch07prb} we have developed a general theory for computing the whole spectrum of a many-spin particle, including core and surface anisotropy, exchange and dipolar interactions, and taking into account the particle's shape, size, and the underlying crystal structure. We then apply this theory to investigate the effects of surface anisotropy on the resonance field and linewidth.
As has been done in Refs.~\onlinecite{garkac03prl, kacbon06prb, yanesetal07prb} for the static properties (hysteresis cycles, energyscape, and magnetic structure), a relationship is being established on the dynamic level, and in particular in what concerns the FMR characteristics, between the many-spin approach and the effective EOSP approach employed here.
This should also help establish the limit of validity of the EOSP model.
\acknowledgements
We acknowledge financial support from the French-Portuguese Pessoa Programme.
%
%\bibliography{hkbib}
%

%
\end{document}